\documentclass[lettersize,journal]{IEEEtran}

\usepackage{amsmath,amsfonts}
\usepackage{algorithm}
\usepackage{algpseudocode}
\usepackage{graphicx}
\usepackage{subcaption}
\usepackage{listings}
\usepackage{multirow}
\usepackage{xcolor}
\usepackage{xspace}
\usepackage{parcolumns}
\usepackage{wrapfig}
\usepackage{booktabs}
\usepackage{makecell}
\usepackage{adjustbox}
\usepackage{balance}

\newcommand{\FIG}{Fig.}

\AtBeginDocument{%
}

\begin{document}
\title{Cloud Performance Decomposition for Long-Term Performance Engineering: A Case Study}

\author{Shimul~Debnath,
        William~Hart,
        Lori~Pollock,
        Donald~Lien,
        and~Wei~Wang%
\thanks{Shimul Debnath, Donald Lien, and Wei Wang are with the University of Texas at San Antonio, USA. 
E-mail: shimul.debnath@utsa.edu, don.lien@utsa.edu, wei.wang@utsa.edu.}%
\thanks{William Hart and Lori Pollock are with the University of Delaware, USA. 
E-mail: wilhart@udel.edu, pollock@udel.edu.}%
\thanks{This work has been submitted to IEEE for possible publication.}%
}

\maketitle

\begin{abstract}
Cloud performance fluctuates due to factors such as resource contention and workload changes. These factors can be short-term, seasonal, or long-term. Their effects are often intertwined in performance traces, making performance management difficult. Prior work on cloud performance engineering used time-series decomposition to separate these factors. However, existing approaches rely on basic decomposition methods that may miss key variation patterns and fail on traces with complex or intermittent patterns, limiting their usefulness across diverse cloud deployments.

To address this limitation, we propose two time-series decomposition techniques for cloud performance engineering: a hybrid/manual method and a fully automatic method. Through a case study of 11 serverless functions, we show that both approaches can successfully and consistently reveal trends and seasonal cycles, such as weekly and quarterly patterns, which are otherwise obscured. 

As an evaluation and application of the decomposition, we used the decomposed components to predict future performance, yielding mean absolute percentage error (MAPE) values of only 1.8\% (hybrid) and 2.1\% (automatic), significantly outperforming basic time-series methods and deep learning. We further show that decomposition insights can guide practical resource allocation. Using decomposition-informed scaling on AWS, we reduced latency variability by over 60\% and maximum latency by 10\%. Similar experiments on benchmarks on AWS confirmed that seasonal patterns and performance gains generalize beyond our case study. Notably, our findings demonstrate that even a single performance trace contains rich actionable information for guiding cloud management decisions.
\end{abstract}


\section{Introduction}\label{sec:introduction}
Cloud-native applications (CNAs) are highly scalable applications designed for dynamic cloud environments~\cite{CloudNativeDefinitionCNCF,CloudNativeDefinitionAWS}. Their high scalability has made them the preferred choice for hosting large-scale web services~\cite{2018-Varghese-FGCS-ServerlessHPC,2016-Villamizar-CCGRID-ServerlessCost} and machine learning models~\cite{2023-Li-OSDI-AplaServe,2021-Wang-SoCC-CloudNativeModelServing,2024-Lu-arXiv-AINative}, which often demand millisecond-level latencies~\cite{2020-Gujarati-OSDI-DNNServing}. Thus, \textit{performance engineering}—including debugging, optimization, monitoring, and analysis—is critical for CNAs both in development and production~\cite{2019-Janes-ISSREW-Microservice,2022-Fattah-TSC-PerfPred,2017-Jayathilaka-WWW-PerfMon-Cloud-Web}.

However, managing cloud performance is notoriously challenging due to significant performance fluctuations~\cite{2018-Maricq-OSDI,2018-Figiela-CCPE-PerfEval,2017-Thalheim-Middleware-Sieve,2017-Bugbee-SADM-PowerPred}.

\begin{figure}
  \centering
  \includegraphics[width=0.49\textwidth]{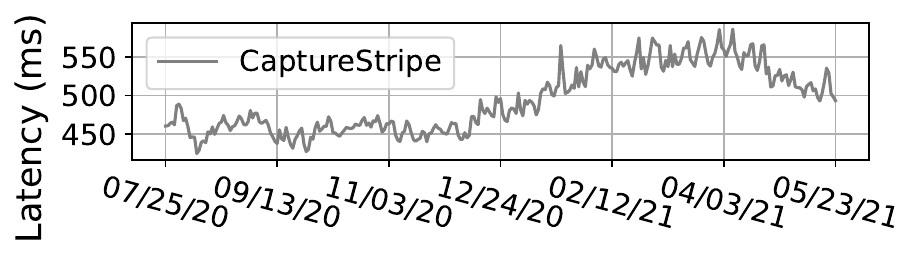}
  \caption{Latency trace of a cloud function over 301 days
    (07/2020$\sim$05/2021) on Amazon AWS Cloud.}
  \label{fig:motivate_trace}
\end{figure}

For example, \FIG~\ref{fig:motivate_trace} illustrates the latency trace of \textit{CaptureStripe}, a serverless CNA from the Serverless Airline Booking Application (SAB), recorded over 43 weeks (301 days) under a constant workload and resource configuration~\cite{EISMANN2022111294}. Despite its stable setup, the application's latency increased substantially from 450ms to 550ms over the year. Such a drastic performance shift necessitates investigation and mitigation, both of which require identifying the cause of the latency increase. Since neither workload nor resource allocation changed, this degradation was likely driven by other factors, such as hardware or software updates in the data center or resource contention due to multi-tenancy. 

However, identifying root causes from performance traces without clear patterns (like the trace in \FIG~\ref{fig:motivate_trace}) is extremely difficult. A \textbf{major challenge} is that cloud performance fluctuations are typically driven by multiple factors acting simultaneously. Their effects intertwine within a single trace, making it difficult to isolate individual effect for reliable analysis. Thus, a \textbf{key requirement} in cloud performance engineering is the ability to \textbf{decompose} a trace into distinct components, each representing a coherent variation pattern.

Prior work has applied time-series decomposition to break cloud performance traces into components such as long-term trends, seasonal (temporal) cycles (e.g., quarterly or monthly), and other variations. For example, isolating long-term trends is widely used in the IT industry to detect performance regressions caused by code updates.~\cite{2014-Vallis-HotCloud-Anomaly,2021-Sarikaa-ICCCA-STLResource,2023-Jia-TETC-SoftwareAging}.

While time-series decomposition is effective for detecting cloud performance regressions, simple techniques used in prior work, such as STL (Seasonal-Trend Decomposition using LOESS)~\cite{STL}, can not always accurately or fully identify the components within cloud performance traces~\cite{2022-Schmidl-VLDBE-AnomalySurvey}, 
especially from public clouds, due to the following challenges.

1. Simple decomposition techniques, such as STL, report all cyclic variations as a single component, without distinguishing between weekly, monthly, or quarterly cycles. While this may suffice for performance regression detection, where only the trend is needed, it is often inadequate for other performance and resource management tasks.

2. Simple decompositions rely on predefined models with fixed parameters. For example, STL applies only LOESS regression with a fixed seasonal period (e.g., weekly). Such fixed-model approaches are less effective for public cloud traces, which often exhibit diverse characteristics (some even non-parametric) and varying periodicities.

3. Cloud performance variations are often non-stationary and intermittent due to the random multi-tenancy. For instance, a weekly cycle (e.g., high latency on Wednesdays) may vanish during holidays. Such irregularity, known as signal \textit{intermittency}~\cite{WEERAKODY2021161,10.1007/978-3-031-70359-1_25}, causes \textit{mode mixing} in simple decomposition methods, where one component includes multiple variation patterns~\cite{4566821,xu2016study}.

To address these challenges, this paper explores advanced time-series analysis techniques for decomposing cloud performance traces. Specifically, we propose two decomposition methods: one hybrid/manual and one fully automatic.

Our hybrid/manual decomposition follows a top-down approach, where users iteratively identify components, starting with the long-term trend and then seasonal patterns. The process continues til the residual appears random, ensuring all significant components extracted. Different model types, parametric or non-parametric, can be applied as needed (thus hybrid).  Users may also fine-tune model parameters to reduce the impact of signal intermittency on automatic fitting.

For users without statistical expertise or want fully automation, we provide an automatic decomposition technique. After evaluating several methods, we found Ensemble Empirical Mode Decomposition (EEMD)~\cite{EMD-Huang,EEMD-WU} to be effective for cloud performance traces. EEMD iteratively extracts components until residuals have low variation, ensuring all relevant components are identified. As a data-driven method, EEMD is not limited by predefined models and is specifically designed to mitigate mode mixing from signal intermittency~\cite{EEMD-WU}.

As a case study, we apply our decomposition techniques to performance traces from the Serverless Airline Booking Application (SAB)~\cite{EISMANN2022111294}, which includes 11 serverless functions in Python, JavaScript, and TypeScript. Amazon developed SAB to showcase CNA designs using cloud-native services.

This case study confirms the effectiveness of our decomposition techniques, as they consistently extract components corresponding to quarterly, monthly, and weekly cycles in performance traces. These findings suggest that cloud performance exhibits a degree of regularity and predictability, enabling performance and resource management strategies to be aligned with these periodic patterns. The results also show that our hybrid/manual and automatic decomposition techniques are largely equivalent, identifying nearly the same components despite their distinct mathematical foundations.

As an application of our decomposition techniques, we developed performance prediction models for the SAB applications to forecast their performance over the next 28 days. These predictions achieved high accuracy, with average MAPE values of only 1.8\% (hybrid/manual) and 2.1\% (automatic). Besides low error rates, our models also effectively captured latency peaks and valleys across most applications. Moreover, our predictions are more accurate than the commonly used STL decomposition and a neural network model. These highly accurate predictions also validate the correctness of our decomposition methods.

As a second application, we applied our decomposition results to optimize SAB’s resource allocation for better performance and stability. Experiments results showed that decomposition-informed allocations reduced latency standard deviation by 60.2\% and slowest latency by 10.8\%, demonstrating that accurate performance decomposition can yield real benefits for cloud deployments. Additional experiments with other benchmarks on AWS and Google Cloud confirm that seasonal patterns and performance benefits generalize across platforms.

To the best of our knowledge, this work is the first to demonstrate how to reliably and completely extract trends and temporal cycles from noisy and irregular public cloud traces. 
Prior to our study, it was unclear whether meaningful trends and temporal cycles could be fully and correctly extracted, and which decomposition techniques were appropriate for such extraction tasks. Our work is novel as we address these open questions by systematically evaluating decomposition methods and demonstrating how to robustly uncover meaningful patterns, enabling informed and practical cloud performance management.

The key contributions of this paper are as follows:

1. \textbf{Hybrid/Manual Decomposition Technique}. A method for effectively identifying trends and seasonal components within cloud performance traces.

2. \textbf{Automatic Decomposition Technique}. A fully automated approach for decomposing cloud performance traces without human intervention, which provides decomposition results nearly equivalent to the hybrid method.

3. \textbf{Comprehensive Case Study}. An analysis of diverse cloud traces, revealing that {\em seemingly random and irregular traces still exhibit regularity and predictability, enabling more effective performance and resource management in the cloud}.

4. \textbf{Two applications of decomposition} – Decomposition-based predictions and resource optimization that not only serve as a practical application of our decomposition techniques but also validate their correctness and benefits.

The rest of this paper is organized as follows:
Section~\ref{sec:decomposition_background} and Section~\ref{sec:background_saba} introduce time-series decomposition and
the SAB serverless application;
Section~\ref{sec:manual_decomposition} and Section~\ref{sec:auto_decomposition} present the hybrid and automatic decomposition techniques, which are compared in Section~\ref{sec:comparison};  
Section~\ref{sec:remaining_functions} evaluates both decomposition
techniques using other SAB functions;
Section~\ref{sec:application} presents decomposition-informed resource optimization;
Section~\ref{sec:related} presents related work;
Section~\ref{sec:discussion} discusses limitations;
and Section~\ref{sec:conclusion} concludes this paper.


\section{Time-series Decomposition Methods}\label{sec:decomposition_background}
Time-series decomposition is a statistical method that breaks a time series into distinct components. As shown in equation~(\ref{eq:time_series})~\cite{Timeseries-Textbook-Chatfield,Brockwell-Davis-TimeseriesText}, a series typically consists of a trend (systematic changes in the mean), seasonal cycles (e.g., quarterly or monthly patterns), other cyclic variations, and random noise. Note that, depending on the data, these components may combine multiplicatively rather than additively in Eq~(\ref{eq:time_series}).
\begin{equation}\label{eq:time_series}
    time\_series = trend + seasonalities + cycles + noise
\end{equation}

A key characteristic of a time series is ``\textit{stationarity}.'' A time series is ``stationary'' if it exhibits no systematic changes in the mean (i.e., no trend) and variance, and has strictly periodic variations removed ~\cite{Timeseries-Textbook-Chatfield}. 
Analyzing non-stationary time series often requires more advanced statistical techniques. 
Due to multi-tenancy, long-term cloud performance data are typically non-stationary, as exemplified by
\FIG~\ref{fig:motivate_trace}.

\subsection{STL Decomposition}\label{sec:stl_decompose}

STL~\cite{STL} decomposition employs
``LOESS smoothing'' 
to extract both trend and seasonal components. 

Due to its simplicity, STL has been widely used in performance anomaly detection for cloud and distributed computing~\cite{2014-Vallis-HotCloud-Anomaly,2021-Sarikaa-ICCCA-STLResource,2023-Jia-TETC-SoftwareAging}. However, it has been reported that anomaly detection using STL may underperform compared to other methods~\cite{2022-Schmidl-VLDBE-AnomalySurvey}.

\begin{figure}
    \centering
    \includegraphics[width=0.4\textwidth]{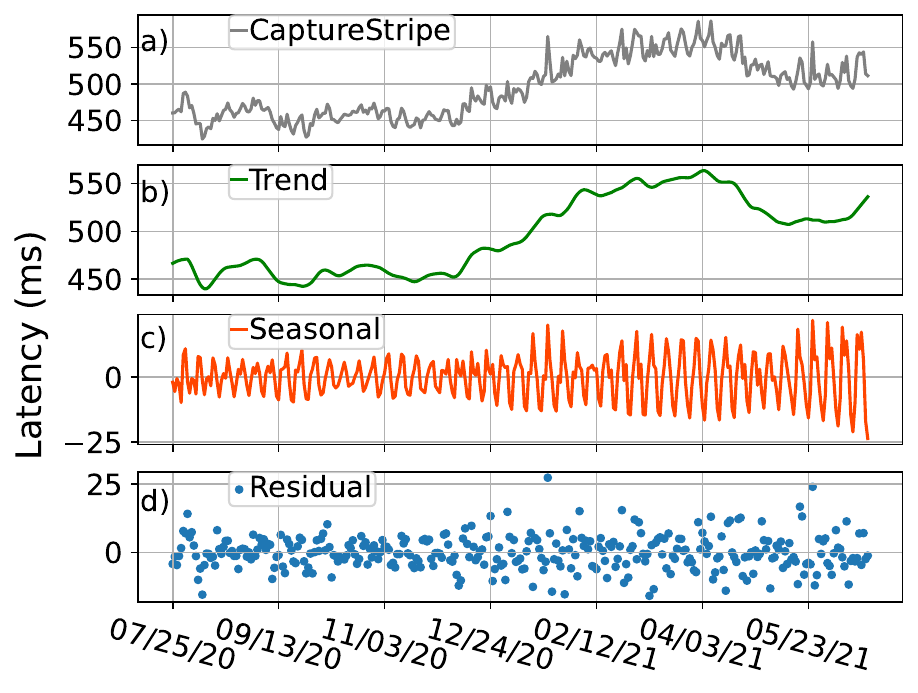}
        \caption{STL decomposition on the trace from
        \FIG~\ref{fig:motivate_trace}.}
        \label{fig:stl_capturestripe}
\end{figure}
\begin{figure}
        \centering
        \includegraphics[width=0.4\textwidth]{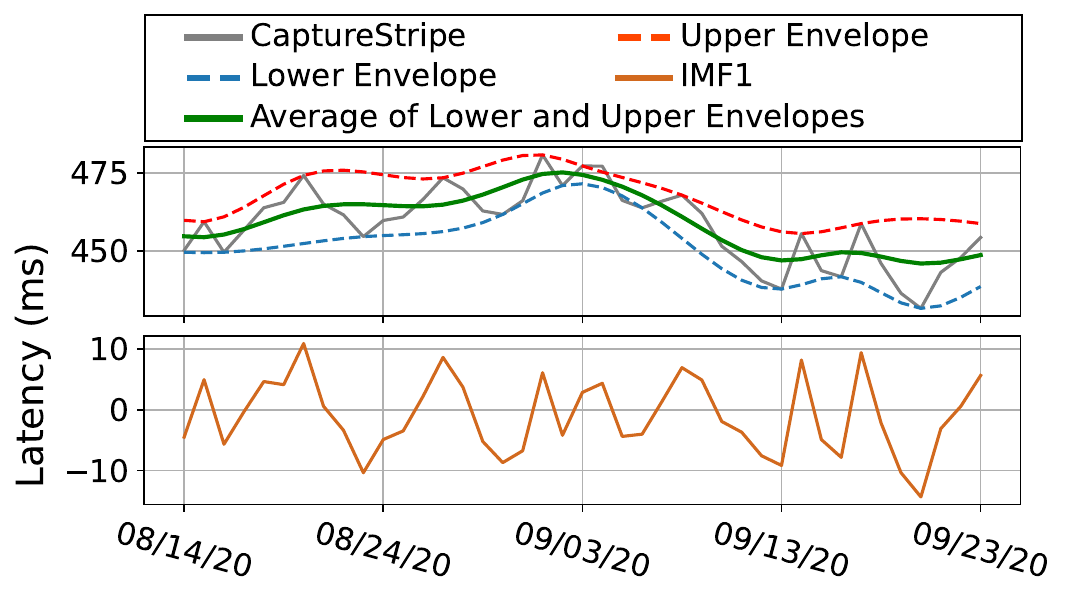}
        \caption{An illustration of one iteration of EMD decomposition using a portion of the trace from \FIG~\ref{fig:motivate_trace}.}
        \label{fig:emd_illustration}
\end{figure}

Our findings align with this report, as we also observed that applying STL naively to cloud performance traces struggles to accurately identify all components. \FIG~\ref{fig:stl_capturestripe} presents the STL decomposition of the performance trace from \FIG~\ref{fig:motivate_trace}. As shown, STL extracts only a single seasonal component with a bi-weekly cycle, while the monthly and quarterly cycles remain undetected. Additionally, a Runs Test~\cite{massoli2024exploring} confirms that the STL residual is not random, indicating the presence of unresolved components.

\subsection{Fourier Transform}
Fourier transform (FT)~\cite{FourierTransform} is a widely used time-series decomposition technique based on frequency analysis. 
FT can decompose a signal into its discrete frequency components, such as distinguishing different sound frequencies in an audio recording. 
However, FT struggles with non-stationary signals~\cite{2023-Eriksen-SciRep-DecompAnalysis}, making it unsuitable for analyzing non-stationary cloud performance traces.

\subsection{Ensemble Empirical Mode Decomposition}
Unlike traditional decomposition techniques, data-driven decomposition imposes minimal prior assumptions, such as stationarity, making them well-suited for handling ``irregular'' time series~\cite{2023-Eriksen-SciRep-DecompAnalysis}. One of the most widely used data-driven methods is Empirical Mode Decomposition (EMD)~\cite{EMD-Huang}, which is purely empirical and places few constraints on the data~\cite{2014-MOTAMEDIFAKHR-EMD,2018-Wang-IEEEAceess-EMD,2020-Stallone-SR-EMD-Assumptions}. The components extracted through EMD are known as \textit{Intrinsic Mode Functions} (IMFs).

EMD is based on the Hilbert-Huang Transform. \FIG~\ref{fig:emd_illustration} illustrates how this transform is applied to the trace from \FIG~\ref{fig:motivate_trace} to generate one IMF (component). The process begins by identifying local maxima and minima in the trace. Next, cubic spline interpolation is used to construct upper and lower envelopes based on these extrema (i.e., the red and blue lines in \FIG~\ref{fig:emd_illustration}). The averages of these envelopes are then computed (shown as the green line in \FIG~\ref{fig:emd_illustration}). Finally, the first IMF (IMF1 in \FIG~\ref{fig:emd_illustration}) is obtained by subtracting the envelope averages from the original trace.

Intuitively, the envelope averages capture the primary fluctuation (variation) in the trace. The difference between the original trace and this primary fluctuation represents the smaller deviations, which form a decomposed component, i.e., an IMF.

After removing an IMF from the trace, the remaining residual can undergo further decomposition using the Hilbert-Huang Transform. By iteratively applying this process, additional variation components (IMFs) are extracted until the final residual reaches a low standard deviation~\cite{EMD-Huang}.

A key limitation of EMD is the aforementioned ``mode mixing'' issue, where a single IMF may contain multiple small and/or intermittent components~\cite{EEMD-WU,1999-Huang-ARFM-Hibert}. To address this issue, Ensemble EMD (EEMD) was introduced~\cite{EEMD-WU}. EEMD enhances EMD by adding white noise to the original trace, perturbing the signal to facilitate the identification of small intermittent variations.

Beyond EMD/EEMD, other data-driven decomposition techniques exist~\cite{2023-Eriksen-SciRep-DecompAnalysis}. As EEMD performed the best among them, we adopted it in this study.

\subsection{Hybrid/manual Decomposition}
Manually decomposing a trace by selecting different, more suitable mathematical models for each component is also common~\cite{2018-Hyndman-ForecastingBook,2022-Lv-AE-HybridDecomp,2016-Oliveira-Neurocomputing-HybridDecomp,2003-Zhang-Neurocomputing-HybridModel}. For instance, trend identification can be performed using methods like moving average, linear regression, piecewise linear regression, or LOESS regression~\cite{2018-Hyndman-ForecastingBook}. Seasonal components can be extracted using various cyclic models, such as sinusoidal regression, ARIMA~\cite{2018-Hyndman-ForecastingBook}, or exponential smoothing~\cite{ExponentialSmoothing}. This hybrid modeling approach is particularly effective for decomposing cloud performance traces, which exhibit diverse behaviors.

Furthermore, hybrid decomposition 
also allows adjusting of model parameters. For instance, if a trace has intermittent weekly cycles, determining the cycle period can be difficult for automatic fitting due to this intermittency. However, a hybrid decomposition can use a cyclic model with a fixed 7-day period to model the weekly cycles. By manually tuning parameters, the decomposition becomes more robust against signal intermittency, reducing the risk of skewed results.

The process of selecting models and their parameters also enhances the reasoning of performance fluctuations. For example, if a decomposed component successfully captures a weekly cycle, it clearly indicates that the performance fluctuation is driven by the differences among weekdays. 

This need for explainability is also why we avoid using neural networks (NN) in our decomposition.


\section{Case Study Data Set}\label{sec:background_saba}
As a case study, we applied our decomposition techniques to the year-long performance traces from the Serverless Airline Booking Application (SAB)~\cite{EISMANN2022111294}. Developed by AWS, SAB serves as a representative cloud-native application (CNA), showcasing best practices for building CNAs using AWS services. It provides comprehensive coverage of AWS cloud-native services, such as cloud-native databases (DynamoDB), event-driven workflows (AWS Step Functions), content delivery (CloudFront), messaging and notifications (AWS SNS), and instrumentation (CloudWatch).

\begin{figure}
  \centering
  \includegraphics[width=0.47\textwidth]{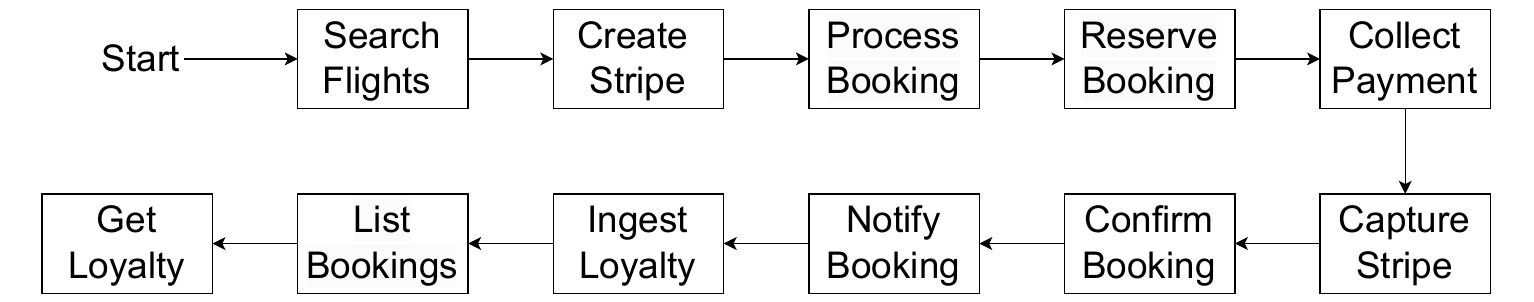}
  \vspace{2mm}
  \caption{The invocation chain of the SAB functions.}
  \label{fig:saba_workflow}
\end{figure}

\FIG~\ref{fig:saba_workflow} illustrates all 11 functions within SAB along with their invocation chain. These functions encompass a standard flight booking process, including flight searches, booking transactions, and user account management. Note that, a SAB function may be invoked by itself (i.e., out of the chain) depending on the use case.

\begin{figure*}
    \centering
    \includegraphics[width=0.9\textwidth]{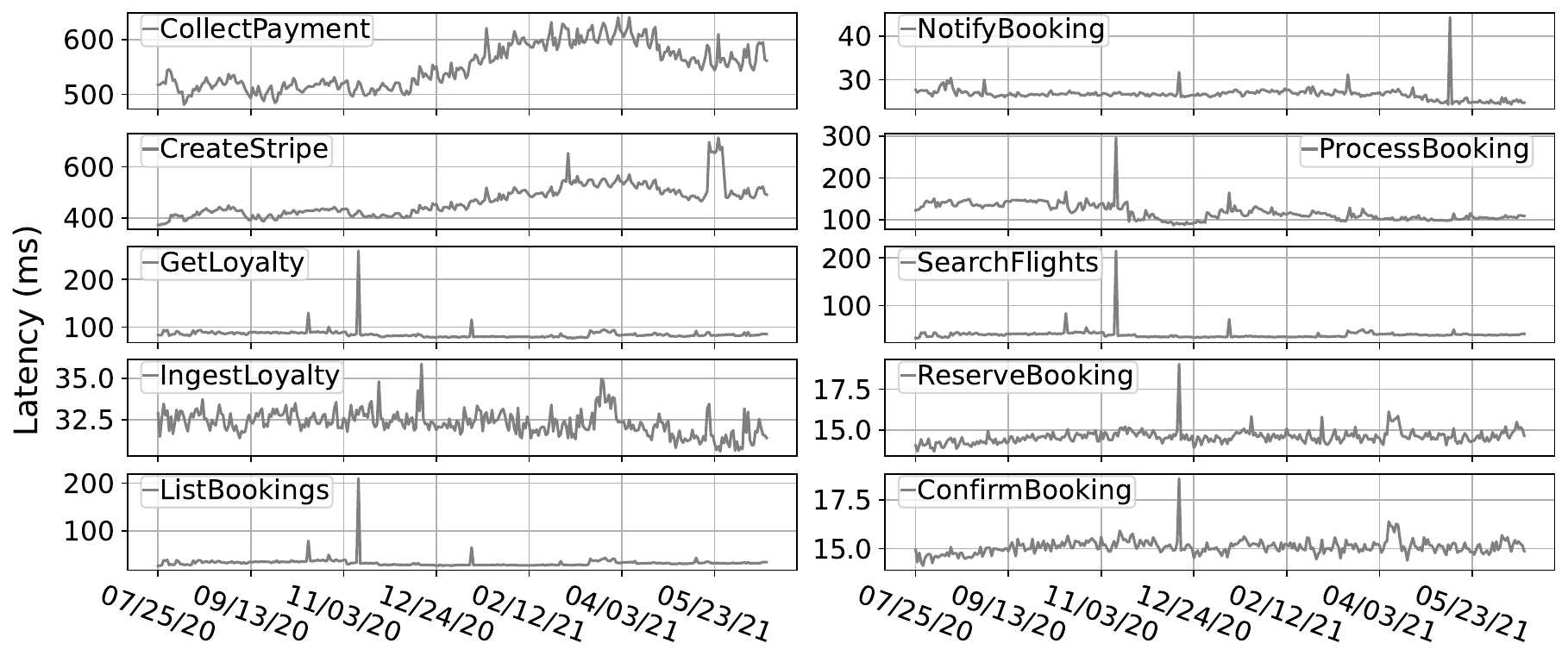}
    \caption{Latency traces of SAB functions (the trace of \textit{CaptureStripe} is in \FIG~\ref{fig:motivate_trace}).}
    \label{fig:saba_traces}
\end{figure*}

\FIG~\ref{fig:saba_traces} presents the performance traces of SAB functions, collected from July 2020 to June 2021. Specifically, for each function, daily performance data under 100 simultaneous invocations were recorded over 329 days (47 weeks). Our decompositions were applied to the first 301 days (43 weeks), while the remaining 28 days (4 weeks) served as test datasets to demonstrate the predictive capabilities of our decomposition, as well as its evaluation.

As shown in \FIG~\ref{fig:saba_traces}, the 11 functions exhibit diverse behaviors and distinct patterns of performance fluctuations. Successfully decomposing all of them highlights the versatility of our techniques.
It is worth noting that some traces may appear similar due to shared underlying services. For instance, {\it CollectPayment}, {\it CreateStripe}, and {\it CaptureStripe} show comparable patterns as they all rely on the payment service from \textit{stripe.com}. Similarly, {\it ConfirmBooking} and {\it ReserveBooking} exhibit similar trends due to their use of the same DynamoDB. However, despite these similarities, each trace retains unique characteristics--such as randomness, outliers, trends, and seasonality--necessitating individual decomposition processes (also as illustrated later by the decomposed components with different periods shown in Table~\ref{tbl:all_periods}).


\section{Hybrid/Manual Decomposition}\label{sec:manual_decomposition}
\subsection{Overview of Hybrid/Manual Decomposition}

\begin{figure}
  \centering
  \includegraphics[height=2.5cm]{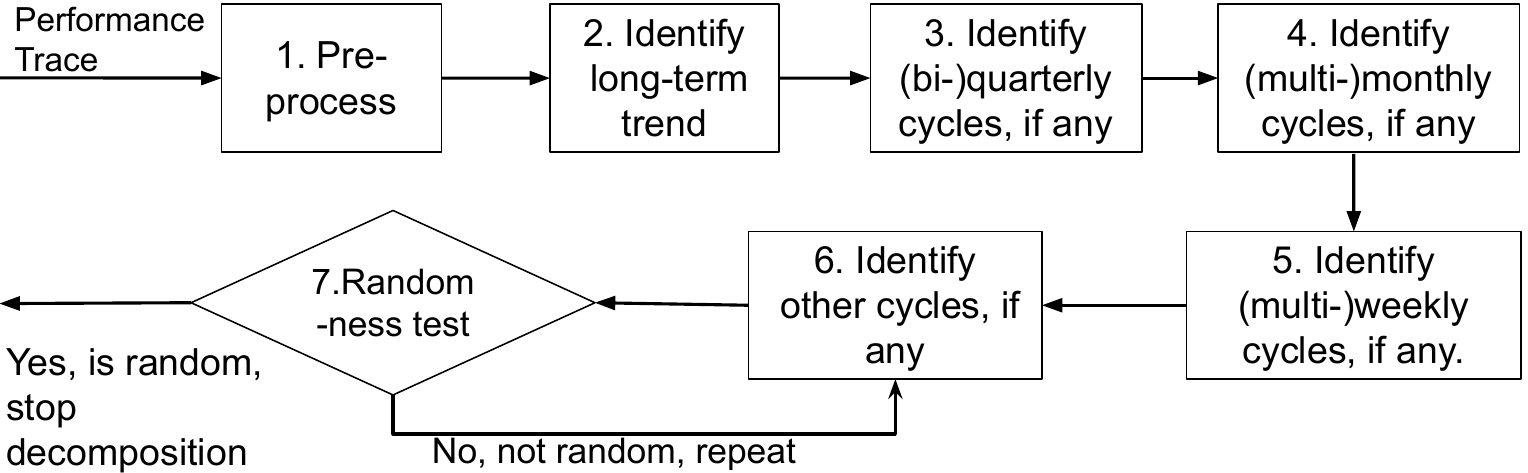}
  \caption{The overall workflow of manual decomposition.}
  \label{fig:manual_decomp_flow}
\end{figure}

\FIG~\ref{fig:manual_decomp_flow} outlines the seven-step workflow of our manual decomposition process. Step 1 processes the input performance trace to remove outliers. Here, we use the Hampel Filter~\cite{pearson2016generalized}, although most outlier-remove algorithms also work. Step 2 involves identifying the long-term trend using a regression model. Users can select a regression model suited to the trace's trajectory. In our case study, we primarily used linear and piece-wise linear regression, though alternatives like moving averages or LOESS can also be applied. Step 3 focuses on detecting quarterly or bi-quarterly cycles, if present. Based on our experience, sinusoidal regression is most effective for these cycles, though any cyclic models can be used, such as ARIMA or Holt-Winters Exponential Smoothing (HWES)~\cite{HWES}. Step 4 identifies (bi-)monthly cycles with cyclic models. Note that, in our case study, we found some traces  lack clear monthly cycles. Hence, cyclic model parameters must be carefully tuned in this step. 
Step 5 captures (bi-)weekly cycles, for which we typically use HWES. Step 6 identifies any remaining cyclic patterns. In our case study, we usually can find semi-weekly cycles with HWES regression. Finally, Step 7 applies the Runs Randomness Test~\cite{massoli2024exploring} to the residual. If the residual is deemed random, the decomposition is complete; otherwise, Step 6 is repeated.

\subsection{Hybrid/Manual Decomposition Example}
\begin{figure*}
    \centering
    \includegraphics[width=0.98\textwidth]{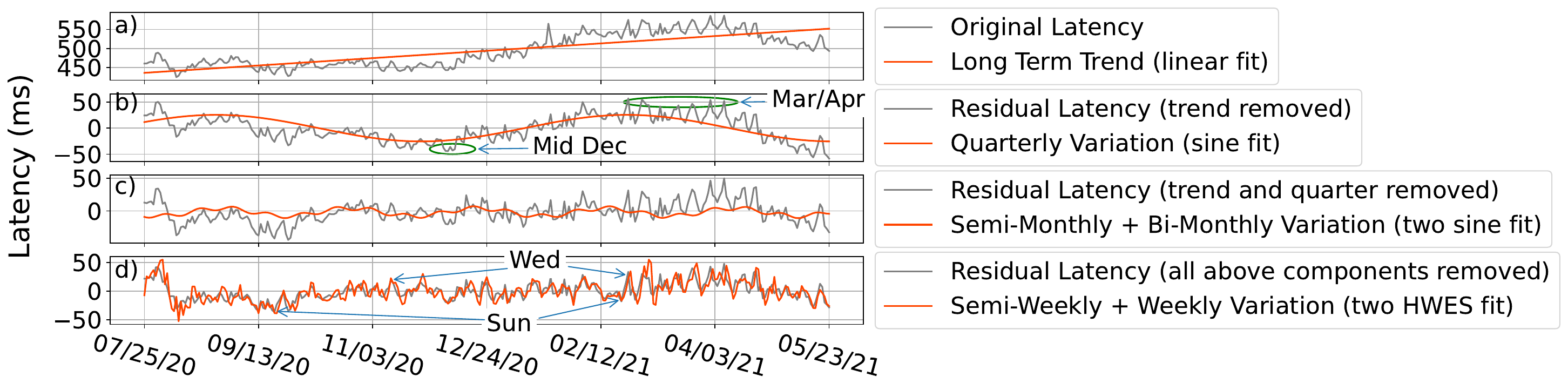}
    \caption{Trend and seasonal variations decomposed from the trace in \FIG~\ref{fig:motivate_trace} using the hybrid decomposition technique.}
    \label{fig:prelim_decomposition}
    \vspace{-2mm}
\end{figure*}

To illustrate the process of the hybrid decomposition, we apply it to the performance trace of the SAB function \textit{CaptureStripe} from \FIG~\ref{fig:motivate_trace}. The decomposition results are presented in \FIG~\ref{fig:prelim_decomposition}, and the detailed cycle periods of each component are given in Table~\ref{tbl:all_periods}.

Following the steps outlined in \FIG~\ref{fig:manual_decomp_flow}, we began by preprocessing and identifying the long-term trend (Steps 1\&2). Noting the steady increase in latency, we hypothesized that the mean latency followed a linear trend~\cite{Timeseries-Textbook-Chatfield}. Consequently, we applied linear regression to model this trend, as shown in \FIG~\ref{fig:prelim_decomposition}a). If this hypothesis holds, the observed linear increase could be attributed to hardware or software changes within the cloud infrastructure~\cite{2020-Uta-NSDI-PerfVar,2018-Gunawi-TS-FailSlow,2018-Figiela-CCPE-PerfEval}. Or, more plausibly, it may result from a progressively increasing background load -- \textit{CaptureStripe} invokes web services from \textit{stripe.com}, an online payment service. It is likely that \textit{stripe.com} experienced a steady rise in workload from 2020~\cite{StripeLoad}, causing \textit{CaptureStripe} to exhibit a corresponding linear increase in latency.

For Step 3, which involves identifying quarterly cycles, we analyzed the residual latency trace after removing the linear trend (i.e., the difference between the original latency and linear regression). As shown in \FIG~\ref{fig:prelim_decomposition}b), the residual latency resembles a sine wave, indicating bi-quarterly variations. The first lowest point appeared around mid-December 2020, aligning with the common expectation that data center usage declines during the holiday season, leading to lower resource contention and improved latency. Observations that align with real-world expectations serve as circumstantial evidence supporting the validity of a decomposition~\cite{Timeseries-Textbook-Chatfield,2021-Ates-ICAPAI-Counterfactual}.

\FIG~\ref{fig:prelim_decomposition}b) also shows a peak around March/April 2021, followed by a decline in May 2021. Seasonal fluctuations in web traffic (high in spring and low in summer) are common~\cite{SeasonalLoad1,SeasonalLoad2,SeasonalLoad3}. 
Higher demand in spring can increase resource contention, resulting in higher latency, while lower demand in summer can reduce contention and improve performance. Given the sine-shaped pattern in \FIG~\ref{fig:prelim_decomposition}b), we fitted a sine function with a 180-day period to represent this bi-quarterly cyclic component.

\FIG~\ref{fig:prelim_decomposition}c) presents the residual latency after removing the bi-quarterly variation. To identify potential monthly patterns (Steps 4), we performed weekly averaging on the residuals, which revealed multiple sinusoidal patterns. Consequently, we applied two sine regressions and identified a bi-monthly component (about 56 days) and a semi-monthly (15-day) component. These two components were combined and are shown in \FIG~\ref{fig:prelim_decomposition}c).

\begin{figure}[ht]
        \centering
        \includegraphics[width=0.45\textwidth]{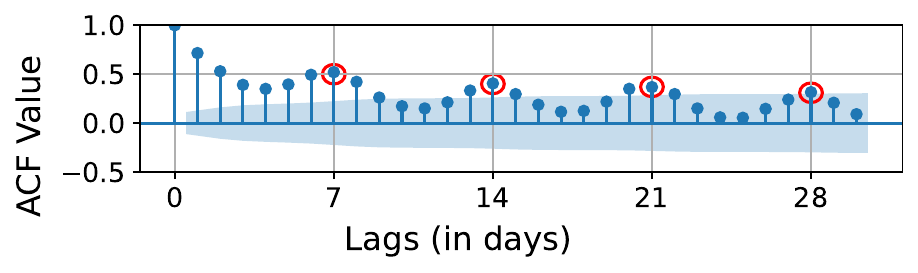}
        \caption{ACF values of the residuals in \FIG~\ref{fig:prelim_decomposition}d). }
        \label{fig:prelim_acf}
\end{figure}
    \begin{figure}[ht]
        \centering
        \includegraphics[width=0.45\textwidth]{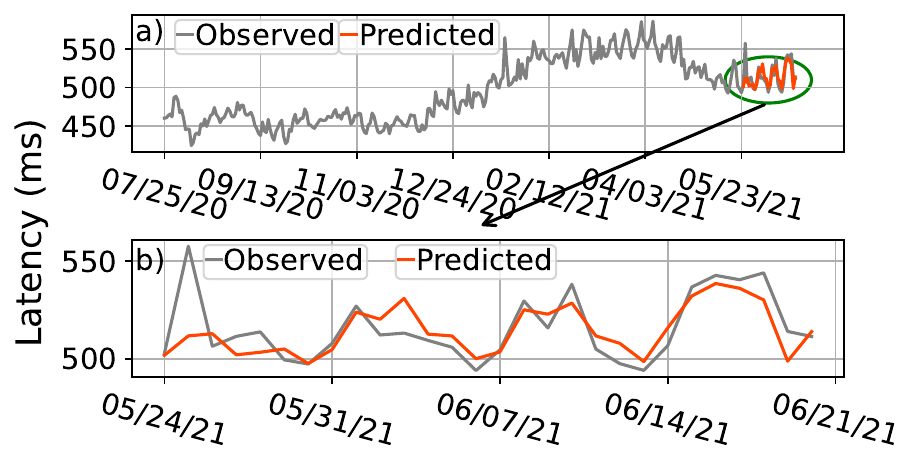}
        \vspace{-2mm}
        \caption{Latency prediction based on the hybrid decomposition for \textit{CaptureStripe}.}
        \label{fig:prelim_prediction}
\end{figure}

\FIG~\ref{fig:prelim_decomposition}d) presents the residues after removing all previously identified components. The residual exhibits clear weekly cycles, with peak latencies typically occurring on Tuesdays, Wednesdays or Thursdays and lowest latencies on Sundays or Saturdays. \FIG~\ref{fig:prelim_acf} displays the autocorrelation function (ACF) of the residual.
ACF measures how much a time series is correlated with itself at different time lags.
\FIG~\ref{fig:prelim_acf} shows the highest correlations are at lags of 7, 14, 21, and 28 days, confirming the presence of weekly cycles. This aligns with the common understanding that data center load and resource contention peak mid-week and decrease over weekend.

To model this weekly pattern, we applied HWES regression, which identified a weekly (7-day) cycle and a semi-weekly (4.6-day) cycle, as shown in \FIG\ref{fig:prelim_decomposition}d). After these two HWES regressions, the final residuals were tested using the Runs test ~\cite{massoli2024exploring} and deemed random, marking the completion of the decomposition process.

Since a common application of decomposition is prediction, the accuracy of a decomposition can be evaluated based on its prediction performance~\cite{Timeseries-Textbook-Chatfield}. Consequently, we combined the decomposed models to forecast the latency for the next 28 days. The predicted latency values and the corresponding observed (ground truth) latency are shown in \FIG~\ref{fig:prelim_prediction}. We used the first 301 data points as ``training data'' to decompose, then predicted the latency for the following 28 days with decomposed models. Note that, none of the 28 days' ground truth latency data were used in the prediction.

\FIG~\ref{fig:prelim_prediction} demonstrates that the predictions are highly accurate, with an error (Mean Absolute Percentage Error, MAPE) of only 1.5\%. More importantly, our predictions successfully captured most peaks and valleys, along with most of the turning points in the observed latency, indicating that the decomposition correctly identified long-term trend and seasonal variations. This high accuracy further confirms the effectiveness of our decomposition technique.


\section{Automatic Decomposition}\label{sec:auto_decomposition}
Our automated decomposition technique is tailored for scenarios that require performance analysis without manual intervention and for users with limited statistical expertise. As previously discussed, this method is built upon EEMD~\cite{EEMD-WU}. 

\begin{figure*}
    \centering
    \includegraphics[width=0.9\textwidth]{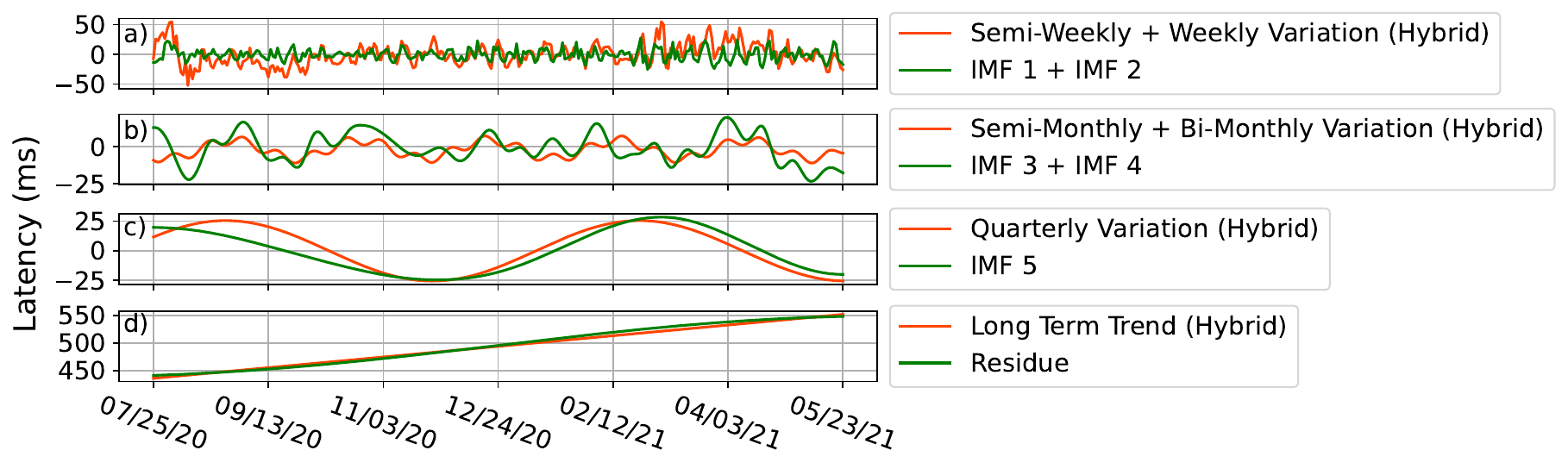
    }
    \caption{Components (IMFs) from the automatic decomposition, with comparisons with the hybrid/manual decomposition.}
    \label{fig:capturestripe_hybrid_vs_automatic_comparison}
\end{figure*}

\subsection{Automatic Decomposition Example}
Again, we demonstrate the automatic EEMD decomposition using \textit{CaptureStripe}'s  trace from \FIG~\ref{fig:motivate_trace}. The decomposition results are presented in \FIG~\ref{fig:capturestripe_hybrid_vs_automatic_comparison}, where EEMD breaks down the trace into six components—five Intrinsic Mode Functions (IMFs) and a residue. Notably, EEMD extracts IMFs in descending order of frequency, with each IMF capturing a variation component with a lower frequency.

Due to space limitation, the first two components, IMF1 and IMF2 are combined in \FIG~\ref{fig:capturestripe_hybrid_vs_automatic_comparison}a).
IMF1 (\FIG~\ref{fig:capturestripe_hybrid_vs_automatic_comparison}a)), captures the most oscillatory component of the trace, characterized by semi-weekly cycles with a 3.3-day period on average. 

Similar semi-week components were found for all traces in our case study with both decomposition methods, suggesting this semi-week fluctuation is a common behavior in AWS cloud.

The second component, IMF2 (\FIG~\ref{fig:capturestripe_hybrid_vs_automatic_comparison}a)) has an average period of 7.6 days. Most of the peaks in IMF2 occur on Tuesdays, Wednesdays, or Thursdays, while most valleys are observed on Saturdays or Sundays. That is, IMF2 captures weekly cycles. 

The third component, IMF3 (\FIG~\ref{fig:capturestripe_hybrid_vs_automatic_comparison}b)), corresponds to semi-monthly  cycles with an average period of 18.3 days, while the fourth component, IMF4 (\FIG~\ref{fig:capturestripe_hybrid_vs_automatic_comparison}b)), captures bi-monthly cycles with an average period of 54 days. These semi-monthly and bi-monthly patterns were also identified by the hybrid decomposition. However, the cycles in IMF3 and IMF4 appear less regular than their counterparts from the hybrid approach, particularly around Nov. and Dec. 2020, where the semi-monthly variations are less visible in the IMF3. This absence of cycles is a case of signal intermittency, which is captured by intermittency-aware EEMD (and the hybrid/manual decomposition).

IMF5 (\FIG~\ref{fig:capturestripe_hybrid_vs_automatic_comparison}c)) shows notable lows during December and peaks around March, represents the bi-quarterly seasonal fluctuation, which was also found by the hybrid decomposition. 
The last component, residual (\FIG~\ref{fig:capturestripe_hybrid_vs_automatic_comparison}d)), reveals a nearly linearly increasing trend, reflecting consistent growth in latency over time, similar to the linear trend found by the hybrid decomposition.

\begin{figure}
    \centering
    \includegraphics[width=0.45\textwidth]{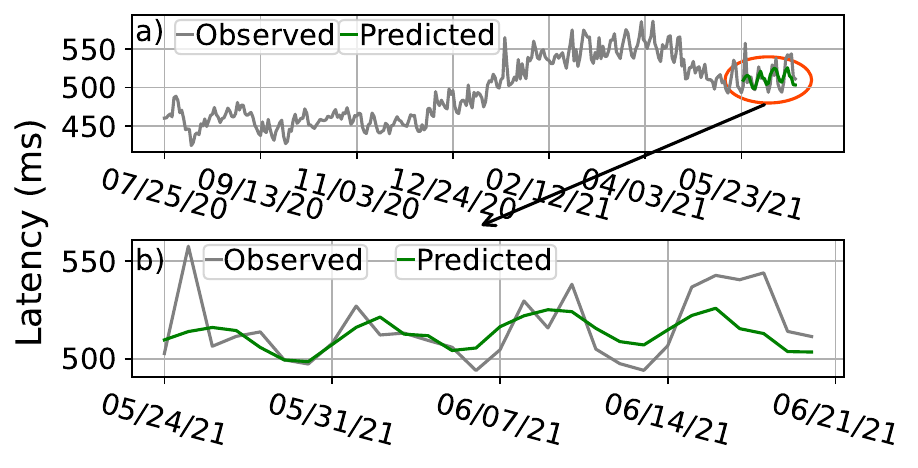}
    \vspace{-2mm}
    \caption{Prediction based on Automatic decomposition for \textit{CaptureStripe}.}
    \label{fig:capturestripe_decomposition_prediction_EEMD}
\end{figure}

Similar to the hybrid decomposition, we also use the automatic decomposition results to predict the latency for the next 28 days. 

\FIG~\ref{fig:capturestripe_decomposition_prediction_EEMD} illustrates the predicted latency alongside the observed latency. The prediction achieves a low error (MAPE) of just 2.0\%. 
The prediction also correctly captures most peaks and valleys with high fidelity. This precise alignment between predicted and observed values also 

corroborates that our automatic decomposition effectively identifies the underlying trends and seasonal patterns.


\section{Comparison of the Hybrid/Manual and Automatic Decomposition}\label{sec:comparison}

Although Section~\ref{sec:auto_decomposition} showed that both hybrid and automatic decomposition methods identify similar components, differences in amplitude and phase shifts may still exist. Therefore, this section presents a direct comparison of these two techniques using the performance trace from \FIG~\ref{fig:motivate_trace}. The decomposed components from both methods are illustrated and compared in \FIG~\ref{fig:capturestripe_hybrid_vs_automatic_comparison}.

\FIG~\ref{fig:capturestripe_hybrid_vs_automatic_comparison}d) compares the long-term trend components identified by both decomposition techniques. As shown, both methods produce nearly identical trends, with the only notable differences occurring near the endpoints, where the trend from the automatic technique appears slightly flattened. This flattening may result from a common issue in EEMD known as ``end effects,'' where EEMD may incorrectly decompose near the boundaries of a time series~\cite{EndEffect,EndEffect2}. Nonetheless, to determine whether the difference is really caused by "end effects" or if the trend genuinely flattens toward the ends, additional data would be required.

Figure~\ref{fig:capturestripe_hybrid_vs_automatic_comparison}c) compares the quarterly components identified by both techniques. The two components align closely, with only a slight difference at the start of the curve—once again due to the “end effect.” This minor discrepancy causes the automatic decomposition's quarterly component to have a longer period (221 days) than the hybrid version (180 days).

\FIG~\ref{fig:capturestripe_hybrid_vs_automatic_comparison}b) compares the semi-monthly and bi-monthly components identified by both techniques. Overall, these components exhibit highly similar patterns, except for the period between November and February. This discrepancy is due to signal intermittency, where these monthly cycles diminished during the holiday season. Being non-parametric, EEMD captured this temporary loss of cycles in its IMF, whereas the hybrid approach assumed more consistent bi-weekly and bi-monthly cycles throughout the year.

To better understand a cloud application's performance, accurately capturing cycle loss and intermittency is preferable. However, for performance prediction, intermittency in cyclic patterns can negatively impact modeling accuracy, as seen in the automatic approach's higher prediction error than the hybrid decomposition (2.0\% vs. 1.5\% MAPE). Ideally, decomposition should balance both objectives—preserving intermittency while maintaining high prediction accuracy. Achieving this balance, however, requires further research.

Finally, Figure~\ref{fig:capturestripe_hybrid_vs_automatic_comparison}a) compares the weekly and semi-weekly components identified by both techniques. While their periods are largely similar, discrepancies appear between November and February due to signal intermittency. Additionally, there are differences in amplitude between the hybrid and automatic decomposition results at the beginning of the traces. These amplitude differences stem from ``end effects'' and the accumulated differences from other seasonal components.

\renewcommand\thesubfigure{\roman{subfigure}}
\begin{figure*}[htbp]
    \centering

    \subfloat[ GetLoyalty\label{fig:getloyalty_comparison}]{%
        \includegraphics[width=0.49\textwidth]{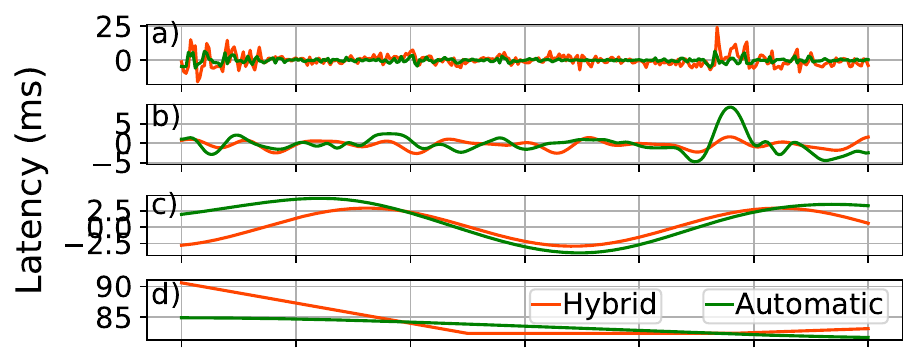}
    }
    \subfloat[ IngestLoyalty\label{fig:ingestloyalty_comparison}]{%
        \includegraphics[width=0.49\textwidth]{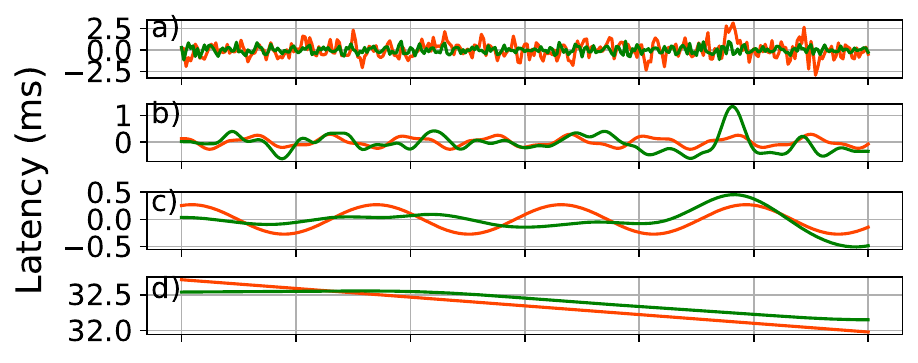}
    }

    \subfloat[ ListBookings\label{fig:listbooking_comparison}]{%
        \includegraphics[width=0.49\textwidth]{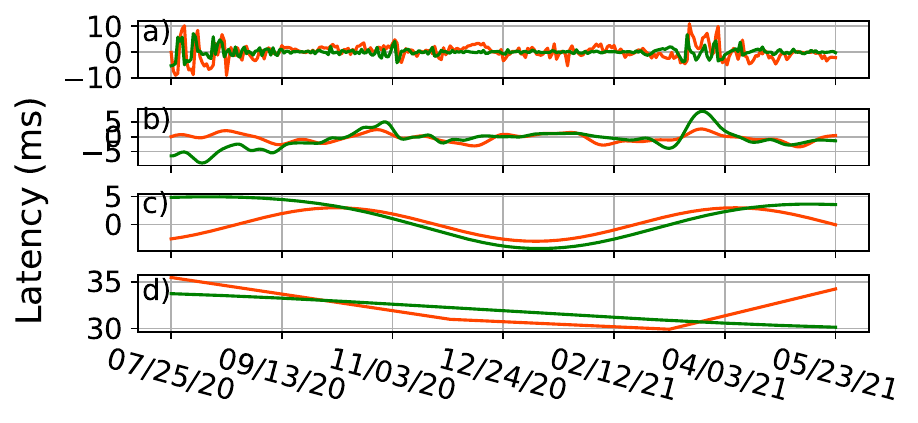}
    }
    \hfill
    \subfloat[ NotifyBooking]{%
        \includegraphics[width=0.49\textwidth]{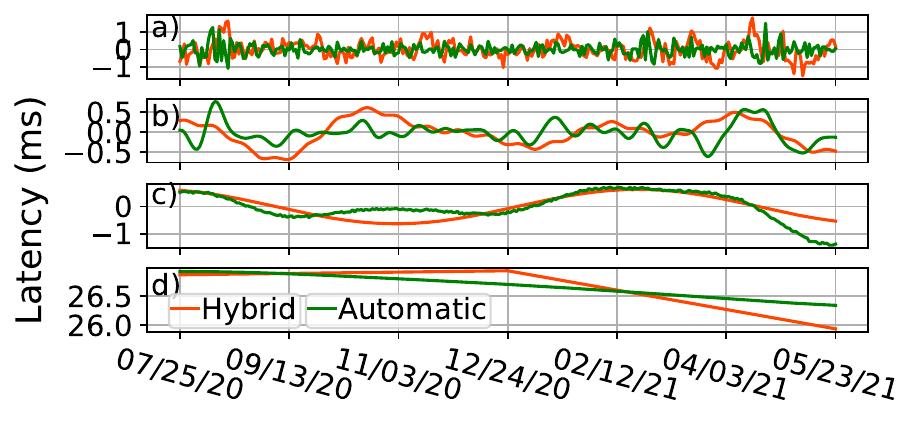}
    }

    \caption{Hybrid and automatic decomposition results for four SAB applications. Each application has four figures representing the summed a) weekly components, b) monthly components, c) quarterly components, and d) long-term trend. }
    \label{fig:additional_comparison}
\end{figure*}

Overall, our hybrid and automatic decomposition techniques yield similar components and show comparable effectiveness across the cloud-native applications studied. Minor differences arise from signal intermittency and ‘end effects,’ which introduce irregular periods and amplitude variations in the automatic results. For anomaly detection, capturing such irregularities can be valuable, whereas for performance prediction, the regularity of the hybrid method is often preferable. In future work, we aim to refine decomposition methods to combine both advantages.


\section{Results of the Complete Case Study}\label{sec:remaining_functions}
This section presents decomposition results for the remaining 10 SAB functions. Table~\ref{tbl:all_periods} lists component periods, while \FIG~\ref{fig:additional_comparison} shows four examples with the largest (though still minor) differences between hybrid and automatic decompositions. Only four are selected due to space constraints.

\subsection{Decomposed Components}

\begin{table*}
\centering
  \fontsize{8}{9}
  \selectfont
\begin{tabular}{|l|l|l|l|l|l|l|l|l|}
\hline
                      & Weekly1 & Weekly2 & Weekly3 & Monthly1 & Monthly2 & Monthly3 & Quarterly & Trend          \\ \hline
CaptureStripe (Hybrid)  & 4.6 & 7 & 15 & 56 & N/A & N/A & 180       & Upward Linear         \\ \hline
CaptureStripe (Auto)  & 3.3 & 7.6 & 18.3 & 54 & N/A & N/A & 221       & Upward Linear    \\ \hline\hline
CollectPayment (Hybrid) & 4.2 & 6.4 & 15 & 56 & N/A & 136 & 209 & Upward Linear         \\ \hline
CollectPayment (Auto) & 3.3 & 7.1 & 17 & 54.2 & N/A & N/A & 205 & Upward Linear    \\ \hline\hline
CreateStripe (Hybrid)   & 4 & 6.5 & 14 & 54.7 & N/A & N/A & 180 & Upward Linear         \\ \hline
CreateStripe (Auto)   & 3.4 & 7 & 15.3 & 44.4 & N/A & N/A & 196 & Upward Linear    \\ \hline\hline
GetLoyalty (Hybrid)     & 3.7 & 5.5 & 21 & 30 & 60 & N/A & 180 & Decline$\rightarrow$Plateau, 3-Piece Lin. \\ \hline
GetLoyalty (Auto)     & 3.1 & 7.2 & 17.1 & 35.6 & 76.5 & N/A & 223 & Downward Linear    \\ \hline\hline
ListBookings (Hybrid)    & 4.5 & 6.1 & 18 & 30 & 73.7 & N/A & 180 & Convex, 3-Piece Linear \\ \hline
ListBookings (Auto)    & 3.2 & 6.5 & 16.8 & 37.4 & 74.5 & N/A & 271 & Downward Linear  \\ \hline\hline
SearchFlight (Hybrid)   & 3.9 & 6.1 & 15 & 30 & 72.3 & N/A & 158 & Convex, 2-Piece Linear \\ \hline
SearchFlight (Auto)   & 3.1 & 6.9 & 15.2 & 30 & 78 & N/A & 156 & Convex Curve \\ \hline\hline
ConfirmBooking (Hybrid) & 4.3 & 5.7 & 16 & 27.5 & 70 & N/A & N/A & Rise$\rightarrow$Plateau, 2-Piece Lin.    \\ \hline
ConfirmBooking (Auto) & 3 & 6.8 & 15.9 & 30.2 & 77 & N/A & N/A &  Rise$\rightarrow$Plateau curve \\ \hline\hline
IngestLoyalty (Hybrid) & 4 & 6 & 15 & 35 & 81 & N/A & N/A & Plateau$\rightarrow$Decline, 2-Piece Lin.               \\ \hline
IngestLoyalty (Auto) & 3.1 & 7.2 & 14.3 & 35.4 & 80 & N/A & N/A & Downward Linear      \\ \hline\hline
ProcessBooking (Hybrid) & 4.5 & 6.1 & 14.3 & 34.9 & N/A & N/A & 126 & Decline$\rightarrow$Plateau, 2-Piece Lin.       \\ \hline
ProcessBooking (Auto) & 3.1 & 7.6 & 16.1 & 39.8 & N/A & N/A & 117 & Decline$\rightarrow$Plateau curve     \\ \hline\hline
ReserveBooking (Hybrid) & 4.1 & 6 & 7 & 28 &  68.7 & N/A & N/A & Rise$\rightarrow$Plateau, 2-Piece Lin.           \\ \hline
ReserveBooking (Auto) & 2.8 & 7.1 & 15.1 & 29.5 & 73.5 & N/A & N/A & Rise$\rightarrow$Plateau curve          \\ \hline\hline
NotifyBooking (Hybrid) & 3 & 6 & 14 & 59.5 & 90 & 129 & 209 & Plateau$\rightarrow$Decline, 2-Piece Lin. \\ \hline
NotifyBooking (Auto) & 3.1 & 6.4 & 14.7 & 61 & 95.5 & N/A & 202 & Downward Linear \\ \hline
\end{tabular}
\caption{Periods (in days) of the decomposed components from the hybrid and automatic decomposition.}
\label{tbl:all_periods}
\end{table*}

\subsubsection{Long-term Trend}

As shown in Table~\ref{tbl:all_periods}, these traces exhibit diverse trends, including upward and downward patterns with potential plateaus, and convex curves that decline before rising. This variety of trends indicates that SAB applications represent a wide range of performance behaviors.

Table~\ref{tbl:all_periods} also shows that trends from the hybrid and automatic approaches largely agree. The hybrid method mainly uses linear or piecewise linear regression for better explainability, while the non-parametric automatic method produces curved but roughly linear trends (sometimes segmented with breakpoints). Four functions—\textit{GetLoyalty}, \textit{ListBookings}, \textit{IngestLoyalty}, and \textit{NotifyBooking} exhibit minor differences, likely due to EEMD’s “end-effect,” as shown in \FIG~\ref{fig:additional_comparison}.

\subsubsection{Seasonal Components}\label{sec:all_sab_seasonality}
Table~\ref{tbl:all_periods} also gives the periods (in days) of each decomposed component from both our hybrid and automatic decomposition methods. Despite the drastically different shapes of the performance traces, their decompositions generally follow weekly, monthly, and quarterly cycles. Specifically, all SAB functions exhibit semi-weekly, weekly, and semi-monthly cycles, while most also display monthly or bi-monthly patterns, along with bi-quarterly cycles. Additionally, a few functions show cycles lasting between 80 to 90 days, roughly corresponding to one season.

As mentioned earlier, these periodic variations are likely driven by cyclic changes in the background load (i.e., multi-tenancy) of the cloud data centers. These background fluctuations lead to variations in hardware resource contention, which in turn cause performance fluctuations. Depending on an application's resource usage (e.g., compute-intensive vs. memory-intensive), the impact of this resource contention can vary. As a result, the specific seasonal components of SAB functions differ slightly.

Additionally, nearly all seasonal components identified by both the methods have similar periods, indicating both detect comparable patterns. However, for \textit{CollectPayment} and \textit{NotifyBooking}, the hybrid method revealed a 130‑day cycle that EEMD missed, as it was embedded within bi‑quarterly components of EEMD. This absence of components reduced prediction accuracy, as shown in Table\ref{tab:performance_traces_mape2}, where the hybrid method produced significantly lower errors.

For \textit{GetLoyalty} and \textit{ListBookings}, the quarterly component periods differ between hybrid and automatic decompositions: 180 days for the hybrid method, versus 223 and 271 days for the automatic method. As with \textit{CaptureStripe} in Section~\ref{sec:comparison}, this discrepancy stems from EEMD’s “end effect,” where IMFs flattening near trace boundaries distorts period estimates. As shown in \FIG s~\ref{fig:getloyalty_comparison}-c) and \ref{fig:listbooking_comparison}-c), the automatic/EEMD curves flatten at the ends, lengthening the apparent period. Nonetheless, the central portions align well, confirming both methods still identified the same quarterly component.

Overall, the seasonal patterns observed indicate that cloud application performance follows clear temporal cycles, suggesting management should account for factors such as weekday and time of year. 
Moreover, while performance traces may appear random, time-series decomposition reveals valuable patterns. These findings call for further research into applying modern statistical techniques to cloud traces to support more informed, large-scale system management.

\begin{figure*}
    \centering
    \includegraphics[width=0.98\textwidth]{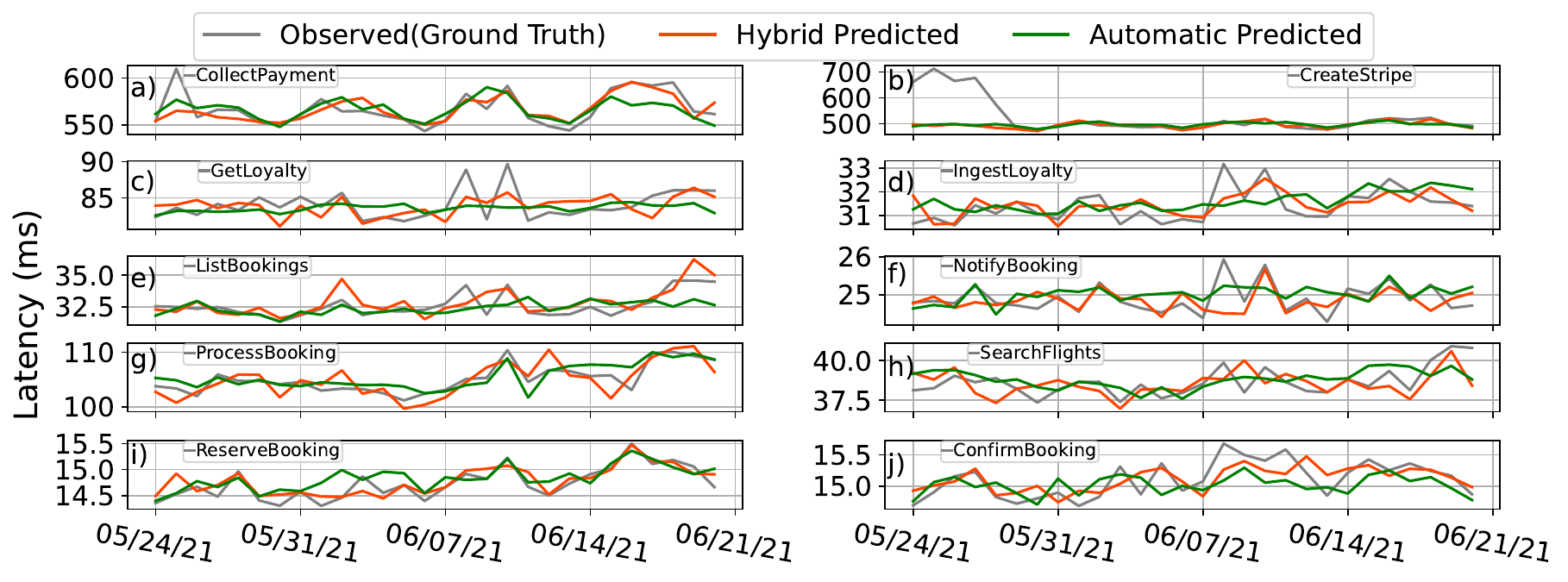}
    \caption{Predicted daily latency of the rest 10 SAB traces. }
    \label{fig:remaining_functions_prediction}
\end{figure*}

\begin{table}[ht]
    \centering
    \fontsize{9}{10}\selectfont
    \setlength{\tabcolsep}{1.5pt}
    \begin{tabular}{l|c|c|c|c}
        \toprule\makecell{\textbf{SAB}\\\textbf{Function}} & \makecell{\textbf{Hybrid}\\(MAPE/ERP)} & \makecell{\textbf{Automatic}\\(MAPE/ERP)} & \makecell{\textbf{LSTM}\\(MAPE/ERP)} & \makecell{\textbf{STL}\\(MAPE/ERP)} \\
        \midrule
        CollectP.    & 1.3\% / 1.3 & 1.7\% / 1.7 & 4.5\% / 4.5 & 4.5\% / 4.5 \\
        CreateS.      & 5.4\% / 3.0 & 6.0\% / 3.2 & 5.9\% / 3.2 & 6.9\% / 3.7 \\
        CaptureS.     & 1.5\% / 1.4 & 2.0\% / 1.9 & 2.9\% / 2.6 & 3.8\% / 3.5 \\
        SearchF.     & 1.8\% / 0.1 & 1.9\% / 0.1 & 2.3\% / 0.1 & 2.4\% / 0.1 \\
        ListB.      & 1.9\% / 0.1 & 2.0\% / 0.1 & 2.4\% / 0.1 & 2.4\% / 0.1 \\
        GetL.        & 1.7\% / 0.2 & 1.7\% / 0.2 & 2.0\% / 0.3 & 4.3\% / 0.6 \\
        ConfirmB.    & 1.1\% / 1.0 & 1.4\% / 1.3 & 2.0\% / 1.9 & 2.5\% / 2.3 \\
        ReserveB.    & 0.8\% / 0.6 & 1.3\% / 0.9 & 1.7\% / 1.3 & 2.6\% / 2.0 \\
        IngestL.     & 1.3\% / 2.1 & 1.9\% / 2.6 & 2.1\% / 3.5 & 2.5\% / 3.9 \\
        NotifyB.     & 1.0\% / 0.3 & 1.6\% / 0.4 & 1.3\% / 0.4 & 2.0\% / 0.6 \\
        ProcessB.    & 1.5\% / 0.2 & 1.1\% /  0.2 & 2.9\% /  0.4 & 5.7\% /  0.8 \\
        \textbf{Average}  & \textbf{1.8\% / 0.9} & \textbf{2.1\% / 1.1} & \textbf{2.7\% / 1.7} & \textbf{3.6\% / 2.0} \\
        \bottomrule
    \end{tabular}
    \caption{Prediction errors (MAPE and ERP) for SAB.}
    \label{tab:performance_traces_mape2}
\end{table}

\subsection{Performance Prediction Results}
\label{sec:prediction_comparison}

We applied both our decompositions to predict SAB’s performance over next 28 days. As shown in Table~\ref{tab:performance_traces_mape2}, both methods achieved low average MAPE errors: 1.8\% for hybrid and 2.1\% for automatic. The hybrid approach performed better, benefiting from human expertise in identifying additional seasonal components, such as the 130-day cycles in \textit{CollectPayment} and \textit{NotifyBooking} (Section~\ref{sec:all_sab_seasonality}).

The largest prediction errors were observed for \textit{CreateStripe} due to the outliers in the groundtruth, which can be seen with large spike at the beginning of \FIG~\ref{fig:remaining_functions_prediction}b). When outliers are removed, the prediction errors are only 1.1\% and 1.7\% for the hybrid and automatic decomposition methods. 

In addition to the low prediction error, the predicted curves from our decomposition methods accurately capture the peaks and valleys of performance fluctuations, as shown in \FIG~\ref{fig:remaining_functions_prediction}. The figure also demonstrates that the hybrid decomposition more closely follows the ground truth curve compared to the automatic approach. Overall, these highly accruate predictions suggest that our decompositions are effective and likely capture all the variation components in cloud performance traces.

For comparison, we also applied STL decomposition and LSTM network to predict SAB latency. Table~\ref{tab:performance_traces_mape2} shows both performed worse than our methods, with STL having the highest average error (3.6\%). Moreover, the curves predicted by STL and LSTM frequently mis-predicted the peaks and valleys. \footnote{For the clarity of the figures, STL and LSTM curves are omitted from \FIG~\ref{fig:remaining_functions_prediction}.}  

They might even predict flat lines, missing all variations. This failure can be quantified using ERP (Edit Distance with Real Penalty)~\cite{ERP}, which measures the shape differences of the predicted and actual time series by computing their normalized distance (lower is better). Table~\ref{tab:performance_traces_mape2} shows that LSTM and STL yield much higher ERP values than our methods, reflecting frequent mispredictions of variations.

Moreover, unlike our decomposition methods, neither STL nor LSTM can fully report seasonal patterns: LSTM offers limited interpretability as a neural network, while STL extracts only a single cyclic component, as shown in \FIG~\ref{fig:stl_capturestripe}.


\section{A Use Case Study: Resource Allocation Based on Decomposition 
Results}\label{sec:application}
\begin{table}[ht]
    \setlength{\tabcolsep}{2.5pt} 
    \centering
    \footnotesize
    \begin{tabular}{lcccc}
        \toprule
        \textbf{Function} & 
        \multicolumn{2}{c}{\textbf{\makecell{Decomposition \\ Informed}}} & 
        \multicolumn{2}{c}{\textbf{\makecell{All-Weekday \\ (naive optimization)}}} \\
        \cmidrule(lr){2-3} \cmidrule(lr){4-5}
        & \makecell{Std. \\ Reduc.} & \makecell{Slowest Latency \\ Reduction} & \makecell{Std. \\ Reduc.} & \makecell{Slowest Latency \\ Reduction} \\
        \midrule
        CollectPayment   & 55.1\% & 6.0\%  & 57.0\% & 6.0\% \\
        CreateStripe     & 46.7\% & 10.3\% & 40.0\% & 10.3\% \\
        CaptureStripe    & 54.0\% & 6.1\%  & 55.4\% & 6.1\% \\
        SearchFlights    & 52.6\% & 12.7\% & 52.6\% & 12.7\% \\
        ListBookings     & 66.7\% & 7.7\%  & 57.8\% & 7.6\%  \\
        GetLoyalty       & 58.5\% & 10.3\% & 64.1\% & 11.5\%   \\
        ConfirmBooking   & 78.3\% & 11.9\% & 87.0\% & 14.1\%  \\
        ReserveBooking   & 66.7\% & 12.1\% & 71.4\% & 12.1\%  \\
        IngestLoyalty    & 73.4\% & 17.2\% & 85.9\% & 20.3\%  \\
        NotifyBooking    & 53.6\% & 7.6\%  & 53.6\% & 7.6\%  \\
        ProcessBooking   & 56.8\% & 10.2\% & 57.7\% & 10.2\%  \\
        \textbf{Average} & \textbf{60.2\%} & \textbf{10.2\%} & \textbf{62.0\%} & \textbf{10.8\%} \\
        \bottomrule
    \end{tabular}
    \caption{Standard deviation and slowest latency reduction of SAB function latencies under the optimized executions.}
    \label{tab:latency_stddev_reduction}
\end{table}

\begin{table}[ht]
  \setlength{\tabcolsep}{3pt}
  \centering
  \footnotesize
  \begin{tabular}{llcccc}
    \toprule
    \textbf{Cloud} & \textbf{Benchmark} &
    \multicolumn{2}{c}{\textbf{\makecell{Decomposition \\ Informed}}} &
    \multicolumn{2}{c}{\textbf{\makecell{All-Weekday \\ (naive optimization)}}} \\
    \cmidrule(lr){3-4}\cmidrule(lr){5-6}
    & & \makecell{Std. \\ Reduc.} & \makecell{Slowest Lat. \\ Reduction} & \makecell{Std. \\ Reduc.} & \makecell{Slowest Lat. \\ Reduction} \\
    \midrule
    \multirow{3}{*}{AWS}
      & IO    & 57.9\% & 6.3\%  &  52.3\% & 6.9\%  \\
      & DB    & 60.0\% & 15.2\% &  70.0\% & 16.6\%  \\
      & \textbf{Average} & \textbf{59.0\%} & \textbf{10.8\%} & \textbf{61.2\%} & \textbf{11.8\%} \\
    \bottomrule
  \end{tabular}
    \caption{Standard deviation and slowest latency reduction of Amazon cloud benchmarks under the optimized executions.}
    \label{tab:latency_stddev_reduction_io_db_aws_google}
\end{table}

\begin{table}[ht]
  \centering
  \footnotesize
    \begin{tabular}{llcc}
      \toprule
      \textbf{Cloud} & \textbf{Benchmark} & \makecell{\textbf{Hybrid}\\(MAPE/ERP)} & \makecell{\textbf{Automatic}\\(MAPE/ERP)} \\
      \midrule
      \multirow{2}{*}{AWS}    & IO & 3.04\%/1.5 & 2.65\%/1.4 \\
                              & DB & 3.17\%/1.6 & 4.24\%/2.2 \\
      \bottomrule
    \end{tabular}%
  \caption{Prediction errors (MAPE and ERP) of Amazon cloud's IO and DB benchmarks}
  \label{tab:decomposition_error_rate_aws_google_io_db}
\end{table}

\subsection{Resource Allocation Optimization for SAB}
Performance decomposition aids not only prediction but also tasks like performance debugging~\cite{2024-Yoon-SOSP-FBDetect} and resource allocation. This section presents a use case applying decomposition insights to optimize cloud resource allocation.

Our decomposition results for SAB in Section~\ref{sec:remaining_functions} show all SAB functions have clear weekly cycles, with higher (slower) latencies on certain weekdays. To improve latency stability and ensure consistent user experience, SAB administrators can choose to optimize resource allocations on the slower weekdays. Therefore, in this use case study, we optimized SAB’s execution on AWS by increasing VM memory allocation to 1024MB memory on each function's slower weekdays (as indicated by the decomposition results),  while maintaining the normal 512MB allocation on the other days.

Each SAB function ran with the adjusted resource allocations for two weeks in August 2025 to evaluate its performance (optimized or decomposition-informed execution). For comparison, another set ran concurrently with normal allocations (unoptimized execution).

\begin{figure*}[!t]
\captionsetup[subfigure]{labelformat=parens,labelsep=space, skip=1pt,belowskip=0pt} 
  \captionsetup{subrefformat=parens, skip=2pt}                
  \renewcommand\thesubfigure{\alph{subfigure}}
  \centering
  \begin{subfigure}{\textwidth}
    \centering
    \includegraphics[width=\textwidth]{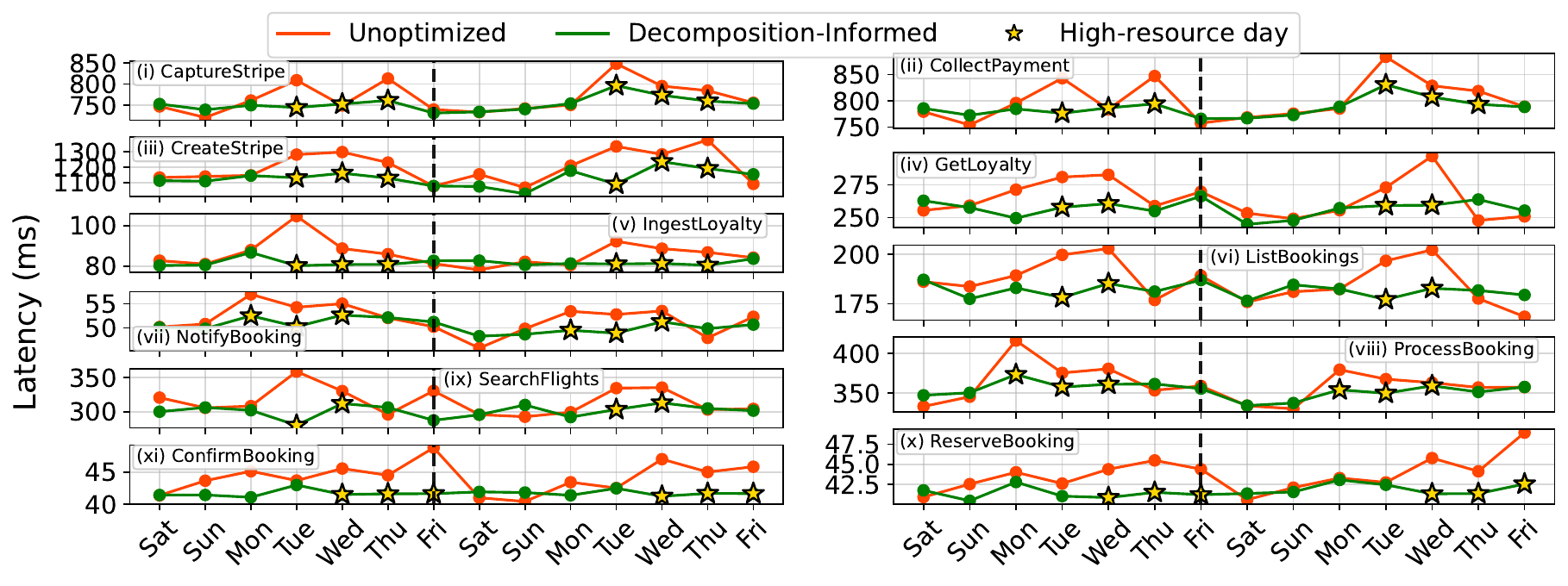}
    \caption{All SAB functions}
    \label{fig:resource_allocation}
  \end{subfigure}

  \vspace{2pt} 

  \begin{subfigure}{\textwidth}
    \centering
    \includegraphics[width=\textwidth]{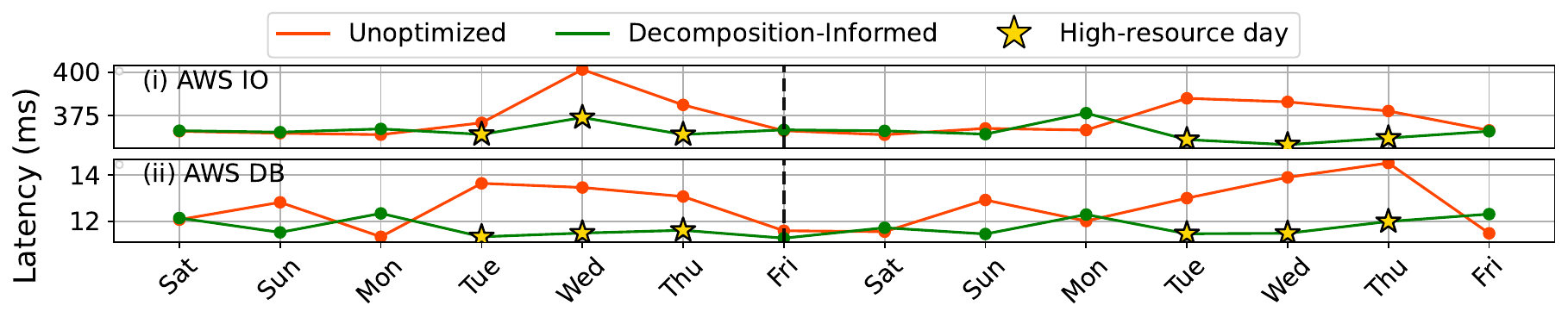}
    \caption{AWS IO and DB benchmarks}
    \label{fig:resource_allocation_io_db_aws_google}
  \end{subfigure}

  \caption{Two-weeks latency under normal unoptimized vs. decomposition-informed allocation.
  Gold-asterisk points indicate the weekdays that were executed with higher resources allocation (i.e., optimized days), while the rest days were not optimized.}
  \label{fig:resource_allocation_combined}
\end{figure*}

\FIG~\ref{fig:resource_allocation} compares latency under optimized and unoptimized executions. Recall that only the slower weekdays identified by decomposition (marked with gold asterisk) were optimized. As \FIG~\ref{fig:resource_allocation} shows, a SAB function typically had two or three slower weekdays, varying from Monday to Friday. \FIG~\ref{fig:resource_allocation} also shows that latency improved substantially on these days for optimized executions. Overall, optimizing the slower weekdays reduced maximum latency by 10.2\% and standard deviation by 60.2\% on average (Table~\ref{tab:latency_stddev_reduction}), greatly improving performance and stability.

As an additional comparison, we also evaluated naively optimizing all five weekdays. As Table~\ref{tab:latency_stddev_reduction} shows, optimizing all weekdays reduced slowest latency by 10.8\% and standard deviation by 62.0\%, similar to the gains from decomposition-informed execution. However, as decomposition-informed execution increased resource usages on fewer days, it achieves similar performance at a lower cloud usage cost. Overall, these results show that the insights from decomposition can indeed translate to real benefits for cloud deployments.

\subsection{Resource Optimization for Other Benchmarks}
To show that our decomposition approach extends beyond SAB, we repeated the weekday-optimization experiments on two SeBS benchmarks~\cite{SeBS}, targeting cloud storage (AWS S3) and database (AWS DynamoDB). The benchmarks were slightly adapted for continuous deployment.

These two benchmarks were executed on AWS Cloud for four weeks, with first two weeks’ data decomposed to identify the slower weekdays. The decomposition revealed clear weekly cycles and potential bi-weekly cycles. Prediction errors are shown in Table~\ref{tab:decomposition_error_rate_aws_google_io_db}. Due to space constraints, we omit detailed decomposition results.

We then ran the benchmarks under both optimized (decom-position-informed) and unoptimized configurations for two more weeks. Results are shown in \FIG~\ref{fig:resource_allocation_io_db_aws_google} and Table~\ref{tab:latency_stddev_reduction_io_db_aws_google}. Decomposition-informed execution reduced slowest latency by 10.8\% and standard deviation by 59.0\%. Table~\ref{tab:latency_stddev_reduction_io_db_aws_google} also shows that the performance gain was also comparable to the naive all-weekday optimization, but at lower cost due to only selected weekdays were optimized. These findings demonstrate that (1) seasonal variations extend beyond SAB, (2) decomposition works across benchmarks, and (3) decomposition insights translate into tangible performance gains across benchmarks.

\section{Related Work}\label{sec:related}

Twitter applied STL decomposition to cloud performance traces to identify long-term trends before detecting performance anomaly~\cite{2014-Vallis-HotCloud-Anomaly}. Similarly, Meta also applied STL decomposition before performance anomaly detection~\cite{2024-Yoon-SOSP-FBDetect}. However, as discussed in Section~\ref{sec:stl_decompose}, STL decomposition cannot completely and correctly decompose performance traces from public clouds. 

Unlike the more controlled environments of internal data centers at Twitter and Meta, public clouds exhibit higher contention and greater randomness, making accurate decomposition more challenging.

Besides decomposition, time-series analysis has been applied in fault detection in the HPC and cloud~\cite{2019-David-SC-FaultHPC,2021-Lima-JS-FaultSurvey,2021-Hagemann-AICCC-CloudAnomalySurvey,2019-Ren-KDD-AnomalyDetect,2015-Ibidunmoye-ACMSurvey-AnomalyDetect}. One work employed ARMA model and Fault Tree Analysis to predict failures~\cite{2012-Chalermarrewong-ICPADS-FaultPred}. Tavakoli et al. treated log data as windowed time series to schedule straggler jobs~\cite{2016-Tavakoli-ICPPW-StragglerSched}. Taerat et al. showed that ARIMA models could predict HPC faults in certain cases~\cite{2010-Taerat-IPDPA-FaultPred}. Gottumukkala et al. used the Weibull distribution to model HPC system reliability from time series data~\cite{2010-Gottumukkala-TR-FacultPred}. Costa et al. used machine learning to cluster similar HPC data to detect application I/O behaviors and potential variations~\cite{2021-Costa-SC-IOVariaton}. Due to the complex nature of fault detection, recent studies tend to employ deep learning rather than just time series analysis~\cite{2022-Schmidl-VLDBE-AnomalySurvey,2019-Ren-KDD-AnomalyDetect}. 

Our work is complementary to these studies as prior work has shown that accurate
performance decomposition can improve the accuracy of fault detection by removing seasonality~\cite{2014-Vallis-HotCloud-Anomaly,2024-Yoon-SOSP-FBDetect}.  Similarly, our decomposition techniques can also improve the accuracy of the root causes identification for cloud and HPC performance issues by isolating the impact of individual factors~\cite{2019-Gan-ASPLOS-Seer,2020-Ma-WWW-AutoMAP,2022-Li-KDD-RootCause,2022-Li-FSE-FaultLocalization,2022-Gan-SIGOPS-Sage,2022-Inagaki-CLOUD-RootCause,2022-Rios-CLOUD-FaultLocalization,2023-Xie-DATE-RootCause,2022-Zhang-ATC-CRISP,2003-Aguilera-SIGOPS-PerfDebug,2004-Cohen-OSDI-PerfDebug,2012-Nagaraj-NSDI-PerfDebug,2022-Ranjitha-CLOUD-PerfDebug,2022-Kong-NSDI-PerfDebug,2022-Huck-ProTools-PerfDebug}.

Another group of closely-related studies involves performance prediction and prediction-based resource management~\cite{2020-Liang-ATC-AutoSys,2017-VanAken-SIGMOD-PerfPred,2018-Li-ATC-Metis,2023-Cheng-TSEM-PerfPred,2021-Eismann-Middleware-Sizeless,2022-Fattah-TSC-PerfPred,2013-Delimitrou-ASPLOS-Paragon}. 
Liang et al. leveraged multiple machine learning models to predict performance for distributed systems~\cite{2023-Liang-NSDI-PerfPred}.
Fu et al. studied ML-based performance prediction of HPC and cloud, and concluded that the "inherent variability in performance that fundamentally limits prediction accuracy"~\cite{2021-Fu-NSDI-PerfPred}.
Cherrypick finds the best cloud resource allocation using performance prediction from Bayesian Optimization~\cite{2017-Omid-NSDI-CherryPick}.

Our work differs from these studies as our primary goal is performance decomposition, with prediction serving as a byproduct or an application. That is, rather than forecasting future performance, our work focuses more on identifying the factors or variations that cause cloud performance changes.

\section{Discussion and Future Work}\label{sec:discussion}
Our results show time-series decomposition reveals trends and seasonal patterns in seemingly random cloud performance traces. However, broader practical adoption requires further research to test its applicability across more cloud applications and longer traces. This section discusses limitations and future directions.

\textbf{Cloud Applications beyond the Case Study}. 
The SAB application in our case study integrates cloud-native services such as databases and workflows and was tested at a reasonable scale (100 concurrent invocations). However, as SAB mainly represents web services, other workloads—like scientific computing or machine learning—may exhibit different performance characteristics. Thus, further research is needed to evaluate time-series decomposition across a wider range of cloud applications.

\textbf{Multi-year Performance Traces}. 
While the seasonal cycles identified in our case study are supported by predictions and alignment with intuitive patterns (e.g., lower data center contention during holidays), longer performance traces could offer further validation—especially if the observed seasonal cycles consistently appear across multiple years.

Moreover, longer performance traces can also alleviate the ``end effect'' of EEMD decomposition, where EEMD may fail to decompose correctly near the boundaries of a time series. With multi-year traces, these boundary issues—particularly at year-end—can be reduced, enabling more accurate analysis of performance trends across year boundaries and offering deeper insight into the differences between the hybrid and automatic decomposition methods.

\textbf{Other Decomposition Techniques}.
Time-series decomposition is a rapidly evolving research area. 
While our hybrid and EEMD-based methods performed well in this case study, cloud applications with different performance characteristics or longer traces may benefit from alternative approaches, such as Variational Mode Decomposition~\cite{VMD} or synchrosqueezed transform~\cite{SynchrosqueezedTransform}. As such, further research should evaluate other decomposition methods, especially when exploring broader applicability of decomposition across other cloud workloads and longer performance traces.

\section{Conclusion}\label{sec:conclusion}
\balance
Time-series decomposition is valuable for cloud performance engineering, but prior STL-based methods struggle with complex public cloud traces. We developed two approaches, hybrid/manual and automatic, and applied them to 11 serverless functions. Both effectively decompose diverse traces, revealing trends and seasonal patterns. Their components enable accurate performance prediction, with average MAPE of 1.8\% (hybrid) and 2.1\% (automatic). Crucially, our work demonstrates that even a single cloud trace holds rich insights for informed cloud management.

\section*{Acknowledgment}
We used OpenAI's ChatGPT to assist with language editing and refinement of this manuscript. All substantive content, analysis, and conclusions are the authors’ own.

\bibliographystyle{IEEEtran}
\bibliography{bibfiles/perf_test.bib,bibfiles/wei.bib, bibfiles/education.bib,bibfiles/cloud_systems.bib}

\end{document}